\documentclass[twocolumn,showpacs,amsmath,amssymb,aps,prd]{revtex4}
\usepackage{latexsym}
\usepackage[dvips]{graphicx}
\begin{document}
\title{Brane-$f(R)$ gravity and dark matter }
\author{A. S. Sefiedgar}
\email{a-sefiedgar@sbu.ac.ir}
\author{Z. Haghani}
\email{z_haghani@sbu.ac.ir}
\author{H. R. Sepangi}
\email{hr-sepangi@sbu.ac.ir}
 \affiliation{Department of Physics,
Shahid Beheshti University, G. C., Evin, Tehran 19839, Iran}
\begin{abstract}
The collision-free Boltzmann equation is used in the context of brane-$f(R)$ gravity to derive the virial theorem. It is shown that the virial mass is proportional to certain geometrical terms appearing in the Einstein field equations and contributes to gravitational energy and that such a geometric mass can be attributed to the virial mass discrepancy in a cluster of galaxies. In addition, the galaxy rotation curves are studied by utilizing the concept of conformal symmetry and notion of conformal Killing symmetry. The field equations may then be obtained in an exact parametric form in terms of the parameter representing the conformal factor. This provides the possibility of studying the behavior of the angular velocity of a test particle moving in a stable circular orbit. The tangential velocity can be derived as a function of the conformal factor and  integration constants, resulting in a constant value at large radial distances.  Relevant phenomenon such as the deflection of light passing through a region where the rotation curves are flat and the radar echo delay are also studied.
\end{abstract}
\maketitle
\vspace{2cm}
%8888888888888888888888888888888888888888888888888888888888888
\section{Introduction}

The question of dark matter is presently one of the most pressing open problems in cosmology.
The galaxy rotation curves and mass discrepancy in a cluster of galaxies are two prominent
observational evidence for the existence of dark matter. According to Newtonian gravity, galaxy rotation curves give the velocity of matter rotating in a spiral disk as a function of the distance from the center of a galaxy; it increases linearly within the galaxy and drops off as the square root of $1/r$  outside the galaxy. However, observations show that this is not the case and the velocity remains approximately constant. This provides for the possible existence of a new invisible matter distributed around galaxies which is known as dark matter \cite{A10-1,A2-12}.

The mass discrepancy of clusters as another evidence for the existence of dark matter can be understood when estimating the total mass of a cluster in two different ways; summing the individual member masses within the cluster leads to a total mass which we shall call $M$. Alternatively,  the virial theorem applied to a cluster would yield an estimate of the cluster mass which we call $M_V$. As it turns out, $M_V$ is nearly $20-30$ times greater than $M$ and this difference is known as the virial mass discrepancy \cite{A10-1,A2-12}. The best way to deal with the above discrepancy, it seems, is to postulate dark matter. There are several candidates for dark matter which can be categorized as baryonic or non-baryonic, or, as hot or cold according to their velocity at the time when galaxies were just starting to form.

In spite of many efforts to postulate various forms of matter as dark matter, there is as yet no non-gravitational evidence for it. Moreover, accelerator and reactor experiments do not support the scenarios in which dark matter emerges. Therefore, one may conclude that Einsteins's gravity may break down at the scale of galaxies \cite{A3}. Thus, to deal with the question of dark matter, modifications to  Einstein field equations \cite{A10-3,A111,A3} have become a flourishing method in recent years. One such modification is the brane-$f(R)$ gravity. The idea that our $4$-dimensional universe might be a $3$-brane embedded in a higher dimensional bulk has its roots in string theory \cite{A333-1}. One of the most successful of such higher dimensional models is the Randall-Sundrum (RS) scenario. In the RS scenario, our $4$-dimensional universe is considered as a brane in a $5$-dimensional bulk with an AdS geometry \cite{A333-2,A333-3}. The RS model has had a great success in explaining the hierarchy problem. Another noticeable effort in this direction is the model proposed by Dvali-Gabadadze-Porrati (DGP) \cite{A333-9}. In this model the bulk space is assumed to be flat but there is an additional $4D$-induced gravity term appearing in the action. A self accelerating phase at late times is predicted for the universe in this model. However, the solutions corresponding to the self-accelerating branch  suffer from ghost instabilities while the normal branch admits solutions which are ghost-free  \cite{prd1,prd2,prd3}. The DGP model predicts modifications to gravity at large distances while the RS-type models modify gravity at small scales. Brane-world scenarios have paved the way for a new understanding of dark matter \cite{A333-7}. In a similar vein,  theories of gravity in which the Einstein-Hilbert (EH) action is replaced with a generic function of $R$, the Ricci scalar, have been quite useful in dealing with the question of dark matter \cite{A10,A16}. It would therefore be of interest to consider a brane-$f(R)$ gravity scenario incorporating both the above ideas to deal with the open questions in gravity.

In addition, there are a number of other motivations for such studies. For example, since brane world scenarios with a generic $f(R)$ in a bulk with a single extra dimension can be described by a real scalar field, brane-$f(R)$ models have been studied within the framework of scalar-tensor type theories. Such theories have been used to investigate the stabilization of the distance between the  branes in the context of the RSI model \cite{ata5} and  the problem of the cosmological constant \cite{ata6}. On the other hand, brane-$f(R)$ type theories have been used to answer questions such as dark energy and  the present accelerating phase of the universe \cite{ata}. The present study could be considered as an attempt to narrow the gap in our understanding of one of the problems facing us in the ever increasing complexities of the inner workings of our universe.

Generally speaking, one may start with a generic action involving $f(R)$ without a cosmological constant in  a RS scenario \cite{A333} in conjunction with the virial theorem in  an attempt to account for dark matter. In doing so, the virial theorem together with observational data relating to the velocity of each member of a cluster, provides an estimate  of the mean density and leads to a prediction of the total possible mass. The virial theorem has also been used in model  $f(R)$ theories with a cosmological constant \cite{A10-10,A10-11}, within the context of brane-world scenarios \cite{A10-12,MH}, using both the metric \cite{A16} and  Palatini formalisms \cite{A10}. In this paper we  generalize  the virial theorem in a brane-$f(R)$ scenario  using the collisionless Boltzmann equation. In this process, extra terms will appear in the generalized  virial theorem originating from the modified action in the bulk. The extra terms can be interpreted as a geometric mass and attributed to the mass discrepancy in clusters.

According to recent observations, the tangential velocity of matter moving around the center of a galaxy tends to a constant value as one moves away from the center of the galaxy. Such rotation curves have been studied in the context of  brane-world scenarios by using the concept of conformal symmetry \cite{A3}. We present similar conclusions in a brane-$f(R)$ model by using the idea of conformal Killing symmetry. In this regard,  the  spacetime is assumed to have, in addition to being static and spherically symmetric, a conformal symmetry. If the vector field $\xi$ is the generator of such conformal symmetry, then the spacetime metric $h$ is mapped conformally onto itself along the trajectories of $ \xi $
\begin{align}\label{0}
{\mathcal{L}}_{\xi} h_{\mu \nu}= \psi h_{\mu \nu},
\end{align}
with ${\mathcal{L}}$ being the Lie derivative  and $ \psi $ is the conformal factor.
There is a systematic method of searching for exact solutions, initiated by Herrera and co-workers \cite{z1,z2,z3}, where
the field equations can be obtained in an exact parametric form, with the conformal factor taken as a parameter. This would provide the possibility of studying the behavior of the angular velocity of a test particle moving in a stable circular orbit. The tangential velocity can be derived as a function of the conformal factor and some integration constants. At large radial distances, we obtain a constant value for the tangential velocity by taking  suitable integration constants.  The bending of light is another issue worth  studying in this region \cite{A111,A222}. The computation of the deflection of photons passing through a region of flat rotation curves and their time delay  are useful tools for testing the alternative theories of gravity, an example of which is presented in this work.

%*************************************************************************
\section{Generalized virial theorem in brane-$f(R)$ gravity}
In order to derive the virial theorem for galaxy clusters, it is necessary to introduce
the brane-$f(R)$ model we are interested in. We also need to know the Boltzmann equation
governing the evolution of the distribution function in cluster of galaxies.  By taking
the cluster of galaxies as a system of identical and collisionless point particles, we
utilize the relativistic Boltzmann equation together with the field equations to find the
generalized virial theorem.
%******************************************************************
\subsection{The brane-$f(R)$ gravity}
In a brane-$f(R)$ model, the $5$-dimensional bulk action is taken as
\begin{align}\label{1}
S=\int {d^5x \sqrt{-g} [ f(R)+{\mathcal{L}}_m ] },
\end{align}
where ${\mathcal{L}}_m$ is the matter lagrangian, $g$ is the bulk metric and $R$ is the bulk Ricci scalar \cite{A333}. Variation of $S$ with respect to the bulk metric $g_{AB}$ yields
\begin{align}\label{1.5-variation}
F(R)& R_{AB}-\frac{1}{2}g_{AB} f(R)+g_{AB} \Box F(R)\nonumber\\
&-\nabla _A \nabla_B F(R)=\kappa_5^2 T_{AB},
\end{align}
where $F(R)=\frac{df(R)}{dR}$. The effective Einstein field equations in the bulk can be written as
\begin{align}\label{1.5-bulkfield}
G_{AB}\equiv R_{AB}-\frac{1}{2}R g_{AB}=T_{AB}^{tot},
\end{align}
where
\begin{align}
T_{AB}^{tot}=&\frac{1}{F(R)}\bigg[ \kappa _ 5^2T_{AB}- \big(\frac{1}{2}R F(R)-\frac{1}{2}f(R)\nonumber\\
&+ \Box F(R) \big)g_{AB}+\nabla_A \nabla_B F(R)  \bigg].
\end{align} \label{1.5-fin}
Using the SMS  procedure \cite{SMS1}, the field equations on the brane are given by
\begin{align}\label{2}
 G_{\mu \nu}=8\pi G_N \tau _{\mu \nu}+\kappa_5^4 \pi_{\mu \nu}+Q_{\mu \nu}-E_{\mu \nu}.
\end{align}
We note that
\begin{align}\label{3}
 G_N=\frac{\kappa_5^4 \lambda}{48 \pi},
\end{align}
where $\lambda$ is the brane tension, $\tau_{\mu \nu}$ is the energy momentum tensor on the brane and
$\pi_{\mu\nu}$ is defined in terms of $\tau_{\mu \nu}$ as
\begin{align}\label{4}
\pi_{\mu \nu}=-\frac{1}{4}\tau_{\mu \alpha} \tau_{\nu} ^\alpha +\frac{1}{12}\tau \tau_{\mu \nu}+\frac{1}{8}h_{\mu \nu}\tau_{\alpha \beta} \tau^{\alpha \beta}-\frac{1}{24}h_{\mu \nu}\tau^2.
\end{align}
The electric part of the Weyl tensor is given by
\begin{align}\label{4.5}
 E_{\mu \nu}={\mathcal{C}}^A_{BCD} n_A n^C h_{\mu}^B h_{\nu}^D,
\end{align}
where $n^A$ is the unit vector normal to the 4-dimensional brane and $h_{AB}=g_{AB}-n_An_B$ is the induced metric on the brane.
Furthermore, we have
\begin{align}\label{5}
Q_{\mu \nu}=&\bigg[\Pi(R)h_{\mu \nu}+\frac{2}{3}\frac{\bigtriangledown_A \bigtriangledown_B F(R)}{F(R)}(h_\mu^Ah_\nu^B \nonumber\\&+n^An^Bh_{\mu \nu})\bigg]_{y=0},
\end{align}
and
\begin{align}\label{6}
\Pi(R)\equiv &-\frac{4}{15}\frac{\Box F(R)}{F(R)}-\frac{1}{10}R\left(\frac{3}{2}+F(R)\right)\nonumber\\
&+\frac{1}{4}f(R)-\frac{2}{5}\Box F(R),
\end{align}
where $y$ is the extra dimension and the brane is located at $y=0$. Now, if we define $Q_{\mu \nu}-E_{\mu \nu}$ as
\begin{align}\label{7}
Q_{\mu \nu}-E_{\mu \nu}=8\pi G_N \mathcal{T}_{\mu \nu},
\end{align}
we obtain the field equations
\begin{align}
 G_{\mu \nu}=8\pi G_N( \tau _{\mu \nu}+ \mathcal{T}_{\mu \nu})+\kappa_5^4 \pi_{\mu \nu},
\end{align}
where $\mathcal{T}_{\mu \nu}$ may act as a new matter source on the brane induced by the $f(R)$ action  in the bulk. It is convenient to represent this new matter by
\begin{align}\label{8}
\mathcal{T}_\mu^\mu=(-\rho_X,P^r_X,P^\bot_X,P^\bot_X).
\end{align}
Let us now consider an isolated and spherically symmetric cluster
described by the metric
\begin{align}\label{9}
ds^2=-e^{\nu(r)}dt^2+e^{\lambda(r)}dr^2+r^2d\theta^2+r^2\sin^2\theta
d\varphi^2,
\end{align}
living on the brane.
Suppose that the clusters are constructed from galaxies which are acting as identical and
collisionless particles and described by the distribution function $f_B$. The energy momentum tensor may be
written in terms of $f_B$ as \cite{A10-15}
\begin{align}\label{10}
\tau_{\mu\nu}=\int{f_B m u_\mu u_\nu du},
\end{align}
where $m$ is the cluster's member mass, $u$ is the four velocity of
the galaxy and $du=\frac{du_r du_\theta du_\varphi}{u_t}$ is the
invariant volume element of the velocity space. The energy momentum
tensor of the matter in a cluster is given by an
effective density $\rho_{eff}$ and an effective anisotropic
pressure, with radial $p^{r}_{eff}$ and tangential $p^{\bot}_{eff}$
components \cite{A10-10}. In other words, we have
\begin{align}\label{11}
\rho_{eff}&=\rho\langle u_t^2\rangle ,\qquad
P_{eff}^{r}=\rho\langle u_r^2\rangle ,  \nonumber\\
 &P_{eff}^{\bot}=\rho\langle u_\theta^2\rangle =\rho\langle
 u_\varphi^2\rangle.
\end{align}
Using $\tau_{\mu}^{\nu}=diag(-\rho_{eff},P_{eff}^r,P_{eff}^{\bot},P_{eff}^{\bot})$ and
$u^{\mu}u_{\mu}=-1$, the field equations become
\begin{align}\label{12}
&e^{-\lambda}\left(-\frac{\lambda '}{r}+\frac{1}{r^2}\right)-\frac{1}{r^2}=-8\pi G_N\rho_{eff} +\frac{\kappa_5^4}{12}\bigg[-(\rho_{eff})^2 \nonumber\\&+(P_{eff}^r)^2-2(P_{eff}^r)(P_{eff}^{\bot})+(P_{eff}^{\bot})^2\bigg]-8\pi G_N \rho_X,
\end{align}
\begin{align}\label{13}
&e^{-\lambda}\left(\frac{\nu '}{r}+\frac{1}{r^2}\right)-\frac{1}{r^2}=8\pi G_N P_{eff}^r+\frac{\kappa_5^4}{12}\bigg[(\rho_{eff})^2 \nonumber\\ &+2(P_{eff}^\bot)(\rho_{eff})-(P_{eff}^r)^2+(P_{eff}^{\bot})^2\bigg]+8\pi G_N P_X^r,
\end{align}
and
\begin{align}\label{14}
&e^{-\lambda}\left(\frac{\nu '}{2r}-\frac{\lambda '}{2r}-\frac{\nu' \lambda'}{4}+\frac{\nu''}{2}+\frac{\nu'^2}{4}\right)=8\pi G_N P_{eff}^\bot\nonumber \\&+\frac{\kappa_5^4}{12}\big[(\rho_{eff})^2+(\rho_{eff})(P_{eff}^r)+(P_{eff}^r)^2\nonumber \\ &+(\rho_{eff})(P_{eff}^{\bot})-(P_{eff}^r)(P_{eff}^{\bot})\big]+8\pi G_N P_X^\bot.
\end{align}
We note that the $(\theta \theta)$ and $(\varphi \varphi)$ components of the field
equations are similar.
%************************************************************************
\subsection{The generalized virial theorem}
To derive the virial theorem, we need the Boltzmann equation which
governs the evolution of the distribution function. By integrating this
equation on the velocity space and using the gravitational
field equations, the virial theorem can be obtained. We consider the cluster as an
isolated spherically symmetric system described by
equation (\ref{9}). Furthermore, we assume that the galaxies in the cluster behave like
identical, collisionless point particles. The distribution function
is denoted by $f_B$ which obeys the general relativistic Boltzmann
equation.

In many applications, it is convenient to work in
an appropriate orthonormal frame or tetrad $e^a_\mu(x)$,
where $g^{\mu\nu}e^a_\mu
e^b_\nu=\eta^{ab}$. In the case of the spherically symmetric
line element given by equation (\ref{9}), we introduce the frame of orthonormal vectors as \cite{A16,A10-15}
\begin{align}\label{17}
e^0_\mu=e^{\nu/2}\delta^0_\mu,\      \
e^1_\mu=e^{\lambda/2}\delta^1_\mu,\         \
e^2_\mu=r\delta^2_\mu,\              \
e^3_\mu=r\sin\theta\delta^3_\mu.
\end{align}
We also note that the tetrad components of $u^\mu$ can be written as $u^a=u^\mu
e^a_\mu$. In tetrad components, the relativistic Boltzmann equation can be written as
\begin{align}\label{18}
u^a e^\mu_a \frac{\partial f_B}{\partial x^\mu}+\gamma^i_{b c}u^b
u^c \frac{\partial f_B}{\partial u^i}=0,
\end{align}
where $f_B=f_B(x^\mu,u^a)$ is the distribution function  and
$\gamma^a_{b c}=e^a_{\mu;\nu}e^\mu_b e^\nu_c $ are the Ricci
rotation coefficients \cite{A16,A10-10,A10-15}. We may assume that $f_B$
depends only on the radial coordinate $r$. Thus, the relativistic Boltzmann equation becomes
\begin{align}\label{19}
&u_1\frac{\partial f_B}{\partial
r}-\left(\frac{1}{2}u_0^2\frac{\partial \nu}{\partial
r}-\frac{u_2^2+u_3^2}{r}\right)\frac{\partial f_B}{\partial u_1}
-\frac{1}{r}u_1\left(u_2\frac{\partial f_B}{\partial
u_2}\right.\nonumber \\ &\left.+u_3\frac{\partial f_B}{\partial u_3}\right)
-\frac{1}{r}u_3 e^{\lambda/2} \cot\theta \left(u_2\frac{\partial f_B}{\partial u_3}-
u_3\frac{\partial f_B}{\partial u_2}\right)=0.
\end{align}
The term proportional to $\cot\theta$ must be zero since we have assumed the system to be spherically symmetric. Let us take
\begin{align}\label{20}
u_0=u_t,\    \ u_1=u_r,\    \ u_2= u_\theta,\    \ u_3=u_\varphi.
\end{align}
Multiplying equation (\ref{19}) by $mu_rdu$, integrating over the
velocity space and assuming that $f_B$ vanishes sufficiently rapidly
as the velocities tend to $\pm\infty$, we find
\begin{align}\label{21}
\frac{\partial}{ \partial r}&\left[\rho \langle u_r^2\rangle
\right]+\frac{1}{2}\frac{\partial \nu}{\partial r}\rho \left[\langle
u_t^2\rangle +\langle u_r^2\rangle \right]\nonumber\\
&-\frac{1}{r}\rho\left[\langle u_\theta^2\rangle +\langle
u_\varphi^2\rangle \right]+
\frac{2}{r}\rho\langle u_r^2\rangle =0.
\end{align}
Now, it is
helpful to multiply equation (\ref{21}) by $4\pi r^2$ and integrate
over the cluster volume to obtain
\begin{align}\label{22}
\int_0^R \rho &\left[\langle u_r^2\rangle +\langle u_\theta^2\rangle
+\langle u_\varphi^2\rangle \right]4\pi r^2 dr \nonumber\\
&-\frac{1}{2}\int_0^R\rho\left[\langle u_t^2\rangle +\langle
u_r^2\rangle \right]\frac{\partial \nu}{\partial r}4\pi r^3
dr=0,
\end{align}
where $R$ is the radius of the cluster.
A useful equation is obtained by summing all none zero components of the field equations (\ref{12}-\ref{14})
\begin{align}\label{23}
e^{-\lambda}&\left(\frac{2\nu '}{r}-\frac{\nu'\lambda'}{2}+\nu''+\frac{\nu'^2}{2}\right)=8\pi G_N  \rho
\big[ \langle u_t^2\rangle +\langle u_r^2\rangle  \nonumber \\ &+\langle u_{\theta}^2\rangle +\langle u_{\varphi}^2\rangle \big]+ \rho^2 \frac{\kappa_5^4}{6}  \big[ 2{\langle u_t^2\rangle}^2+2{\langle u_t^2\rangle}{\langle u_{\theta}^2\rangle} \nonumber\\ &+ {\langle u_t^2\rangle}{\langle u_r^2\rangle} \big]+8\pi G_N [\rho_X+P_X^r+2P_X^\bot],
\end{align}
where we have used equation (\ref{11}).
Using $\langle u^2\rangle=\langle u_t^2\rangle+\langle u_r^2\rangle+\langle u_{\theta}^2\rangle+\langle u_{\varphi}^2\rangle$ and assuming that the galaxies in the cluster have velocities
much smaller than the velocity of light \cite{A16,A10-12}, that is $\langle
u_r^2\rangle \approx\langle u_\theta^2\rangle \approx\langle
u_\varphi^2\rangle \ll\langle u_t^2\rangle \approx1$, we obtain
\begin{align}\label{24}
e^{-\lambda}&\left(\frac{2\nu '}{r}-\frac{\nu'\lambda'}{2}+\nu''+\frac{\nu'^2}{2}\right)=8\pi G_N  \rho
  \nonumber \\ &+ \frac{1}{3}\kappa_5^4 \rho^2 +8\pi G_N \left(\rho_X+P_X^r+2P_X^\bot \right).
\end{align}
Since for clusters of galaxies the ratio of the matter density to the brane tension is much smaller than unity, $\frac{\rho}{\lambda} \ll 1$, we can neglect the quadratic term in the matter density in equation (\ref{24}).
We may also define $\bar{\rho}_{X}=\left( \rho_X+P_X^r+2P_X^\bot \right) $, where $\bar{\rho}_X$ is a pure geometric term acting as a new matter source on the brane. It carries all the effects induced on the brane by the $f(R)$ action in the bulk. We therefore have
\begin{align}\label{25}
e^{-\lambda}\left(\frac{2\nu '}{r}-\frac{\nu'\lambda'}{2}+\nu''+\frac{\nu'^2}{2}\right)=8\pi G_N  \rho
+8\pi G_N \bar{\rho}_X.
\end{align}
To move on, we assume that $\nu'$ and $\lambda'$ are slowly varying, i.e. $\nu'$ and $\lambda'$ are small and their product can be neglected, and $e^{-\lambda}\approx 1$ inside the cluster. We then have
\begin{align}\label{26}
\frac{1}{2r^2}\frac{\partial}{\partial r}\left(r^2\frac{\partial \nu}{\partial r}\right)=4\pi G_N  \rho+4\pi G_N \bar{\rho}_{X}.
\end{align}
On the other hand, using the above assumptions, one may write equation (\ref{22}) as
\begin{align}\label{27}
2K-\frac{1}{2}\int_0^R{4 \pi r^3 \rho \frac{\partial \nu}{\partial r}dr}=0,
\end{align}
where
\begin{align}\label{28}
K=\int_0^R{2 \pi \rho \left[\langle u_r^2\rangle+\langle u_{\theta}^2\rangle+\langle u_{\varphi}^2\rangle \right]r^2dr},
\end{align}
is the total kinetic energy of the galaxies.
Multiplying equation (\ref{26}) by $r^2$ and integrating yields
\begin{align}\label{29}
G_NM(r)=\frac{1}{2}r^2\frac{\partial \nu}{\partial r}-G_N M_X(r),
\end{align}
where we have used $M=\int_0^R{dM(r)}=\int_0^R{4\pi\rho r^2dr}$ as the baryonic mass. We have also defined  $M_X=\int_0^R{4\pi\bar{\rho}_X r^2dr}$ as the geometric mass of the system.
Now consider the following definitions
\begin{align}\label{30}
\Omega=-\int_0^R\frac{G_N M(r)}{r}dM(r),
\end{align}
and
\begin{align}\label{31}
\Omega_X=\int_0^R\frac{G_NM_X(r)}{r}dM(r).
\end{align}
Multiplying equation (\ref{29}) by $\frac{dM(r)}{r}$ which is
equal to $\frac{4\pi \rho r^2 dr}{r}$ and integrating gives
\begin{align}\label{32}
\Omega=\Omega_X-\frac{1}{2}\int_0^R 4\pi r^3\rho \frac{\partial
\nu}{\partial r}dr,
\end{align}
where $\Omega$ is the gravitational potential energy of
the system.
Finally, using equation (\ref{27}) leads to the generalized virial theorem
\begin{align}\label{33}
2K+\Omega-\Omega_X=0.
\end{align}
Alternatively, the above equation can be written in the form
\begin{align}\label{33.5}
2K-G_N \int_0^R \frac{M(r)dM}{r}-G_N\int_0^R M_X(r) \frac{dM}{r}=0.
\end{align}
For convenience we introduce the radii $R_V$ and $R_X$
\begin{align}\label{34}
R_V=\frac{M^2}{\int_0^R \frac{M(r)}{r}dM(r)},
\end{align}
and
\begin{align}\label{35}
R_X=\frac{M_X^2}{\int_0^R
\frac{M_X(r)}{r}dM(r)}.
\end{align}
Furthermore, the virial mass $M_V$ is defined as \cite{A16}
\begin{align}\label{36}
2K=\frac{G_N M M_V}{R_V}.
\end{align}
Substitution of these definitions in equation (\ref{33.5}) leads to
\begin{align}\label{37}
\frac{M_V}{M}=1+\frac{M_X^2 R_V}{M^2 R_X}.
\end{align}
From observation, the relation $M_V/M > 3$
holds true for most of the galactic clusters. Therefore one can easily approximate the last equation
\begin{align}\label{38}
\frac{M_V}{M}\approx\frac{M_X^2 R_V}{M^2 R_X}.
\end{align}
Now one may note that some geometric terms appearing in the Einstein field equations above
could effectively play a role in the gravitational energy. These
geometric terms may be attributed to a geometric mass at the
galactic or extra galactic levels and may be interpreted as dark
matter. On the other hand, dark matter is the main contribution of mass in clusters.
It means that the contribution of the baryonic
mass is negligible in comparison with dark matter. The total mass of the cluster can then be estimated as $M_{tot}\approx M_X$. We also know that the virial mass is mainly determined by the geometric mass. It means that the geometric mass
could be a potential candidate for the virial mass discrepancy in
clusters. As a result, we conclude that
\begin{align}\label{39}
M_X \approx M_V\approx M_{tot}.
\end{align}
Therefore, equation (\ref{38}) can be written as
\begin{align}\label{40}
M_V\approx M \frac{R_X}{R_V}.
\end{align}
This shows that the virial mass is proportional to the normal
baryonic mass in a cluster and the proportionality constant has
geometrical origins.

%***************************************************************************************************
\section{Vacuum solution}
In this section, we consider the vacuum solution of the theory. Assuming  $\tau_{\mu\nu}=0 $, the field equations  (\ref{12}-\ref{14}) reduce to
\begin{align}\label{az1}
e^{-\lambda}\left(\frac{\lambda'}{r}+\frac{e^{\lambda}}{r^2}-\frac{1}{r^2}\right)=8\pi G_N \rho_X,
\end{align}
\begin{align}\label{az2}
e^{-\lambda}\left(\frac{\nu'}{r}-\frac{e^{\lambda}}{r^2}+\frac{1}{r^2}\right)=8\pi G_N P_{X}^{r},
\end{align}
and
\begin{align}\label{z2}
e^{-\lambda}\left(\frac{\nu'}{2r}-\frac{\lambda'}{2r}-\frac{\lambda'\nu'}{4}+\frac{\nu''}{2}+\frac{\nu'^{2}}{4}\right)=8\pi G_N P_{X}^{\bot}.
\end{align}
Since the above system of equations contain five unknown quantities, it is under determined. To obtain the unknown quantities we require two extra relations.
Therefore, it is convenient to consider
\begin{align}\label{az3}
\rho_X=P_X^r,
\end{align}
as an equation of state. We also assume that the spacetime admits a conformal group of symmetries \cite{A3}. In other words, if the vector field $\xi$ is the generator of such a conformal symmetry, then the spacetime metric $h$ is mapped conformally onto itself along the trajectories of $ \xi $
\begin{align}\label{z3}
{\mathcal{L}}_{\xi} h_{\mu \nu}= \psi h_{\mu \nu},
\end{align}
where ${\mathcal{L}}$ is the Lie derivative operator and $ \psi $ is the conformal factor \cite{z1,z2,z3}.\
It should be noted that such extra relations affect the form of $f(R)$ through relation (\ref{az3}).
Let us consider a general form for the vector field $\xi$
\begin{align}\label{z4}
\xi=\xi^0(t,r)\frac{\partial}{\partial t}+\xi^1(t,r)\frac{\partial}{\partial r}+\xi^2(\theta,\varphi)\frac{\partial}{\partial \theta}+\xi^3(\theta,\varphi)\frac{\partial}{\partial \varphi}.
\end{align}
Substituting this  conformal vector in equation (\ref{z3}) and assuming  $\psi=\psi(r)$, one obtains
\begin{align}\label{z5}
\xi=\frac{k}{2}t\frac{\partial}{\partial t}+\frac{r\psi}{2}\frac{\partial}{\partial r}+\frac{d F(\varphi)}{d\varphi}\frac{\partial}{\partial \theta}-[\cot(\theta)F(\varphi)-G(\theta)]\frac{\partial}{\partial \varphi},
\end{align}
where $F(\varphi)$ and $G(\theta)$ are arbitrary functions and $k$ is an integration constant \cite{A2}.
This procedure also leads to the metric components in terms of the conformal factor
\begin{align}\label{z6}
e^{\nu(r)}=C^2r^2\exp\left(-2k\int \frac{dr}{r\psi}\right),
\end{align}
and
\begin{align}\label{z7}
e^{\lambda(r)}=\frac{B^2}{\psi^2},
\end{align}
where $B$ and $C$ are integration constants \cite{A2}.
Substitution of equations (\ref{z6}) and (\ref{z7}) into field equations (\ref{az1})-(\ref{z2}) leads to
\begin{align}\label{z8}
\frac{\psi^2}{B^2}\left(\frac{2\psi'}{r\psi}+\frac{1}{r^2}\right)-\frac{1}{r^2}=-8\pi G_N \rho_X,
\end{align}
\begin{align}\label{z9}
\frac{\psi^2}{B^2}\left(\frac{3}{r^2}-\frac{2k}{r^2\psi}\right)-\frac{1}{r^2}=8\pi G_N P_X^r ,
\end{align}
and
\begin{align}\label{z10}
\frac{\psi^2}{B^2}\left(\frac{1}{r^2}+\frac{k^2}{r^2\psi^2}-\frac{2k}{r^2\psi}+\frac{2\psi'}{r\psi}\right)=8\pi G_N P_X^{\bot}.
\end{align}
Using the equation of state (\ref{az3}) and equating equations (\ref{z8}) and (\ref{z9}) we obtain
\begin{align}
r\psi \psi'+ 2\psi^2- k \psi-B^2=0. \label{z11}
\end{align}
This equation can be solved to give $r$ in terms of $\psi$
\begin{align}
r^4=R_0^4\frac{G(\psi)}{|2\psi^2-k\psi-B^2|}, \label{z12}
\end{align}
where $R_0$ is an integration constant and
\begin{align}\label{z13}
G(\psi)=\exp\bigg\{\frac{-2k}{\sqrt{k^2+8 B^2}} \tanh^{-1}\left[\frac{-4\psi+k}{\sqrt{k^2+8 B^2}}\right]\bigg\}.
\end{align}
Now, by substituting all quantities in terms of $\psi$, the energy density and pressure of what may now be called the X-matter can be written in terms of the conformal factor as
\begin{align}\label{z14}
8\pi G_N \rho_X &= 8\pi G_N P_X^r  \nonumber \\ &=\frac{\sqrt{-2\psi^2+k\psi+B^2}}{B^2R_0^2\sqrt{G(\psi)}}\left(3\psi^2-2k\psi-B^2 \right),
\end{align}
and
\begin{align}\label{z15}
8\pi G_N P_X^{\bot}=-\frac{\sqrt{-2\psi^2+k\psi+B^2}}{B^2R_0^2\sqrt{G(\psi)}}\left(3\psi^2-2B^2-k^2 \right).
\end{align}
The sum of equations (\ref{z14}) and (\ref{z15}) is another useful relation
\begin{align}
8\pi G_N(\rho_X + P_X^{\bot})=\frac{1}{B^2r^2}(k^2-2k\psi+B^2),
\end{align}
which can be used to define the equation of state at large distances.
It is clear that in the limit $r\to\infty$, the X-matter has an equation of state $\rho_X=P_X^r=-P_X^{\bot}$.

It should be emphasized that to find the vacuum solution we only ignored ordinary matter on the brane. The presence of $f(R)$ in the bulk action results in modifications which can be traced to $f(R)$-nonlinearity effects emanating from $Q_{\mu \nu}$ and $E_{\mu \nu}$. Such effects constitute what is known as the X-matter. The  vacuum solutions obtained here may thus be usefully utilized to account for the flat rotation curves in a region where the contribution of X-matter is dominant.

%************************************************************************************************************************
\section{Galaxy rotation curves in a conformally symmetric spacetime}
\begin{figure}
\centering
\includegraphics[scale=1]{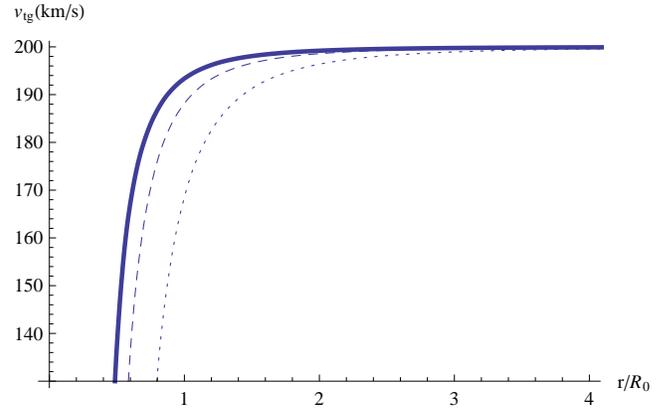}
\caption{Variation of the tangential velocity as a function of $r/{R_0}$, where the dotted line represents $B=1.1000007$,  the dashed line $B=1.1$ and the solid line  $B=1.0999998$.}\label{FIG1}
\end{figure}

The observational data show that the rotational velocity increases linearly within the galaxy and approaches a constant value of about $200 km/s$ as one moves away from the center \cite{A2-12}. In this section, we shall consider a test particle which moves in a circular time-like geodesic orbit and study its tangential velocity. The Lagrangian of the system is given by
\begin{align}
2L&=\left(\frac{ds}{d\tau}\right)^2 \nonumber \\
&=-e^{\nu(r)}\left(\frac{dt}{d\tau}\right)^2+e^{\lambda(r)}\left(\frac{dr}{d\tau}\right)^2+r^2\left(\frac{d\Omega}{d\tau}\right)^2,\label{z20}
\end{align}
where $\tau$ is the affine parameter along the geodesic and $d \Omega^2=d\theta^2+\sin ^2 \theta d\varphi^2$. From the above Lagrangian one obtains
\begin{align}\label{z21}
E=e^{\nu}\dot{t}=\mbox{const}.\qquad  \mbox{and}  \qquad l_{\phi}=r^2\sin^2\theta \dot{\phi}=\mbox{const}.,
\end{align}
where $E$ and $l_{\phi}$ are the energy and $\phi$-component of the angular momentum of the test particle respectively which are conserved quantities  and an over dot represents differentiation with respect to $t$. Although the $\theta$-component of the angular momentum is not a constant of motion, the total angular momentum, $l^2=l_{\theta}^2+(l_{\phi}/\sin\theta)^2$, is a conserved quantity which can be written as $l^2=r^4\dot{\Omega}^2$ \cite{A3-4}.
Equation of the geodesic orbit is given by
\begin{align}\label{z22}
\dot{r}^2+V(r)=0,
\end{align}
where $V(r)$ is the potential
\begin{align}\label{poten}
V(r)=-e^{-\lambda}\left(E^2e^{-\nu}-\frac{l^2}{r^2}-1\right).
\end{align}
Let us now study the motion associated with stable circular orbits. This kind of motion can be obtained by the conditions
\begin{align}\label{az10}
\dot{r}=0 \qquad   \mbox{and}   \qquad \frac{\partial V}{\partial r}=0.
\end{align}
The potential obtained subject to the above conditions describes an extremum of the motion. To have a minimum, the condition
\begin{align}\label{az11}
\frac{\partial^2 V}{\partial r^2} > 0,
\end{align}
is also required.
Using these conditions, one obtains the energy and total angular momentum   \cite{A3-4,A3-5}
\begin{align}\label{z23}
E^2=\frac{2e^{\nu}}{2-r\nu'}\qquad   and     \qquad l^2=\frac{r^3\nu'}{2-r\nu'}.
\end{align}
For an inertial observer far from the source who measures the spatial components of the velocity, normalized to the speed of light, the line element can be rewritten as $ds^2=-dt^2(1-v^2)$ \cite{A3-4}, where
\begin{align}
v^2=e^{-\nu}\bigg[e^{\lambda}\left(\frac{dr}{dt}\right)^2+r^2\left(\frac{d\Omega}{dt}\right)^2\bigg].
\end{align}
Using the condition for circular orbits, the tangential velocity can be expressed as
\begin{align}
v_{tg}^2=r^2e^{-\nu}\left(\frac{d\Omega}{dt}\right)^2.
\end{align}
The tangential velocity can be rewritten in terms of the conserved quantities
\begin{align}
v_{tg}^2=\frac{e^\nu}{r^2}\frac{l^2}{E^2},
\end{align}
or in the following alternative form
\begin{align}\label{az12}
v_{tg}^2=\frac{r\nu'}{2},
\end{align}
where equation (\ref{z23}) is used.
This relation shows explicitly that the tangential velocity depends on the time-time component of the metric only. Previously, we obtained the metric components in terms of the conformal factor in equations (\ref{z6}) and (\ref{z7}). Therefore, using the relation between $\nu$ and $\psi$, the angular velocity in terms of $\psi$ is given by
\begin{align}\label{vtg}
v_{tg}^2=1-\frac{k}{\psi}.
\end{align}
Using equation (\ref{vtg}) and (\ref{z7}), the metric coefficient $\exp(\lambda)$ can also be expressed in terms of the angular velocity
\begin{align}\label{lambda}
\exp(\lambda)=\frac{B^2}{k^2}(1-v_{tg}^2)^2.
\end{align}
Although, we cannot write $v_{tg}$ as a function of $r$ explicitly, it is possible to find $v_{tg}$ in some regions.
For example, we can derive the value of tangential velocity at infinity. It is clear from equation (\ref{z12}) that we have $r\to \infty$ for $\psi= \psi_{1,2}$, where
\begin{align}
\psi_{1,2}=\frac{k}{4}\pm\sqrt{\frac{k^2}{16}+\frac{B^2}{2}}.
\end{align}
Therefore one may obtain the tangential velocity at infinity from equation (\ref{vtg}) as follows
\begin{align}
v_{tg \infty}=\sqrt{1-\frac{k}{\frac{k}{4}\pm\sqrt{\frac{k^2}{16}+\frac{B^2}{2}}}}.
\end{align}
Knowing the value of the tangential velocity at infinity from observations one can derive the relation between $k$ and $B$. The variation of $v_{tg}$ as a function of $r$, using equations (\ref{vtg}, \ref{z11}),   for $v_{tg \infty}=200 km/s$ is shown in Fig.\ref{FIG1}.
It should be noted that parameter $R_0$ is an integration constant from equation (\ref{z12}) and  helps to re-scale the curve only. Furthermore, $B$  should not be looked upon as a fine tuning parameter as it can take  values not necessarily close to $1$. The particular choices for $B$ in Fig.\ref{FIG1} are only made so that the curves are better distinguished.

Now, it is convenient to obtain geometric mass of the galaxy and its dependence on $r$. If we  integrate equation (\ref{az1}), we will find
\begin{align}
e^{-\lambda}=1-\frac{8\pi G_N}{r}\int \rho_X r^2 dr.
\end{align}
Using definition $M_X =4\pi\int_{0}^{r}\rho_X(r')r'^2dr'$, we also obtain
\begin{align}
e^{-\lambda}=1-\frac{2G_N M_X}{r}.\label{lam}
\end{align}
One may now introduce the radius $r=R$ for the vacuum boundary where the contribution of the baryonic mass vanishes, $\rho_b\approx0$.
Of course, the continuity  of the metric coefficients $e^{\lambda}$ and $e^{\nu}$ across the vacuum boundary of the galaxy should be preserved. For simplicity we assume that inside the baryonic matter with density $\rho_ b$ the nonlocal effects of the bulk can be neglected. Therefore at the vacuum boundary the metric coefficients can be written as
\begin{align}\label{lam1}
e^{\nu}=e^{-\lambda}=1-\frac{2G_N M_b}{R},
\end{align}
where $M_b =4\pi\int_{0}^{R}\rho_b(r')r'^2dr'$ is the total baryonic mass of the galaxy. By equating equations (\ref{lambda}) and (\ref{lam1}) at $r=R$, one has
\begin{align}
\frac{k^2}{B^2}=\left(1-\frac{2G_N M_b}{R}\right)\left(1-v_{tg}^2\right)^2.\label{kb}
\end{align}
Using the above  formula, it is possible to find a relation between $M_X$ and $M_b$ from equations (\ref{lambda}) and (\ref{lam}) as follows
\begin{align}
M_X=\frac{M_b}{R}r,
\end{align}
showing that $M_X$ varies linearly with $r$, consistent with observations.

\section{The Lagrangian}
To find the form of the Lagrangian $f(R)$ of the theory on the brane, we need the metric.
Here, we concentrate on the flat rotation curve region for which $v_{tg}$ is constant. According to equation (\ref{vtg}), we have $\psi= k/(1-v_{tg}^2)$. Therefore, the constancy of $v_{tg}$ in the region of flat rotation curves leads to the constancy of $\psi$. As a result we may write
\begin{align}\label{nu}
e^{\nu}=C^2r^{2v_{tg}^2},
\end{align}
and
\begin{align}\label{azlambda}
e^{\lambda}=\frac{B^2}{k^2}\left(1-v_{tg}^2\right)^2,
\end{align}
obtained from equations (\ref{z6}) and (\ref{z7}).
Therefore the form of the line element can be written as
\begin{align}\label{line1}
ds^2=-C^2r^{2v_{tg}^2}dt^2+\frac{B^2}{k^2}(1-v_{tg}^2)^2dr^2+r^2d\Omega^2,
\end{align}
where the relation between $k$ and $B$ is known according to equation (\ref{kb}) and $C$ can be defined according to the continuity of the metric coefficient $e^{\nu}$ across the vacuum boundary of the galaxy. Using equations (\ref{lam1}) and (\ref{nu}) we have
\begin{align}
C^2=R^{-2v_{tg}^2}\left(1-\frac{2G_NM_b}{R}\right).
\end{align}

So far we have found the metric in the region of flat rotation curve as given by (\ref{line1}).
In this region we use the field equations (\ref{2}) on the brane to determine the form of the Lagrangian. By neglecting the contribution of the baryonic matter, the field equations reduce to
\begin{align}
G_{\mu\nu}=Q_{\mu\nu}-E_{\mu\nu}, \label{fe}
\end{align}
where $E_{\mu\nu}$ and $Q_{\mu\nu}$ are given by equations (\ref{4.5}) and (\ref{5}) respectively.  To solve the field equations, we start with the bulk metric and write it as
\begin{align}
ds^2=-N(y)r^{\alpha}dt^2+N(y)Adr^2+r^2d\Omega^2+dy^2 ,\label{bulkline}
\end{align}
where $A$ is a constant and $N(y)$ is a function of the extra dimension. To obtain a consistent line element on the brane as in (\ref{line1}), we assume $\alpha=2v_{tg}^2$, $A=\frac{B^2}{C^2k^2}(1-v_{tg}^2)^2$ and  $N(0)=C^2$. With these assumptions, the metric becomes
\begin{align}
ds^2=-C^2 r^{\alpha}dt^2+\beta dr^2+r^2d\Omega^2, \label{braneline}
\end{align}
where $\beta=C^2A$.
To continue, we can find $f(R)$ from the trace of equation (\ref{1.5-bulkfield}) and substitute it in the brane field equations. The non zero components of the field equations become
\begin{widetext}
\begin{align}\label{Newf0}
\frac{2}{15 \beta r^2}(-4\beta +4-\alpha ^2-\alpha)+\frac{1}{5C^2}\left(\frac{\hat{N}^2}{C^2}-\frac{7\hat{\hat{N}}}{3}\right)+\frac{2}{5\beta r}\frac{F_{,r}}{F}\left(\frac{4}{3}-\alpha\right)+\frac{4}{15\beta }\frac{F_{,rr}}{F}-\frac{1}{5F}\left(2\hat{\hat{F}}+\frac{\hat{N}\hat{F}}{3C^2}\right)=0,
\end{align}
\begin{align}\label{Newf1}
\frac{2}{15 \beta r^2}(-4\beta +4+9\alpha-\alpha^2)+\frac{1}{5C^2}\left(\frac{\hat{N}^2}{C^2}-\frac{7\hat{\hat{N}}}{3}\right)+\frac{4}{15\beta r}\frac{F_{,r}}{F}(\alpha+2)-\frac{2}{5\beta }\frac{F_{,rr}}{F}-\frac{1}{5F}\left(2\hat{\hat{F}}+\frac{\hat{N}\hat{F}}{3C^2}\right)=0,
\end{align}
\begin{align}\label{Newf2}
\frac{2}{15 \beta r^2}(\beta +4\alpha ^2-\alpha-1)-\frac{1}{5C^2}\left(\frac{\hat{N}^2}{4C^2}+\frac{2\hat{\hat{N}}}{3}\right)+\frac{2}{15\beta r}\frac{F_{,r}}{F}(2\alpha-1)+ \frac{4}{15\beta }\frac{F_{,rr}}{F}-\frac{1}{5F}\left(2\hat{\hat{F}}-\frac{4\hat{N}\hat{F}}{3C^2}\right)=0,
\end{align}
where $F=F(r,y)$, $F_{,r}=\frac{\partial F}{\partial r}\mid_{y=0}$ and $F_{,rr}=\frac{\partial^2 F}{\partial r^2}\mid_{y=0}$. In addition, we have defined $\hat{F}=\frac{\partial F}{\partial y}\mid_{y=0}$ and $\hat{\hat{F}}=\frac{\partial^2 F}{\partial y^2}\mid_{y=0}$, with the same notation for $N=N(y)$. We emphasize that $N$ and its derivatives are constant on the brane.
\end{widetext}
From equations (\ref{Newf0}) and (\ref{Newf1}) we find
\begin{align}
-r^2F_{,rr}+\alpha r F_{,r}+2\alpha F=0,
\end{align}
with general solution
\begin{align}\label{FFF}
F(r)=C_1r^{n_1}+C_2r^{n_2},
\end{align}
where $C_1$ and $C_2$ are arbitrary integration constants and $n_{1,2}=(\alpha+1\pm\sqrt{\alpha^2+10\alpha+1})/2$. We limit ourselves to the monotonically decreasing solution so that $F(r)=C_2r^{n}$ where $n= n_2 <0$ \cite{A16}.
Now, we can derive $\hat{F}$ and $\hat{\hat{F}}$ from equations (\ref{Newf0}) and (\ref{Newf2})
with the result
\begin{align}\label{Fhathat}
\hat{F}(r)= C_2r^{n}\left(\frac{f_1}{r^2}+\frac{3\hat{N}}{{4C^2}}-\frac{\hat{\hat{N}}}{\hat{N}}\right),
\end{align}
where $f_{1}=\frac{C^4}{{\beta} \hat{N}}[- \sqrt{\alpha^2+10\alpha+1}(1-\alpha)+3-3\alpha^2-2\beta]$ and
\begin{align}\label{Fhathat}
\hat{\hat{F}}(r)=C_2r^{n}\left(\frac{f_2}{r^2}+\frac{3\hat{N}^2}{{8C^4}}-\frac{\hat{\hat{N}}}{C^2}\right),
\end{align}
with $f_{2}=\frac{1}{{2\beta}}[- \sqrt{\alpha^2+10\alpha+1}+3\alpha-2\beta+3]$.
Using the above formulae and the trace of equation (\ref{1.5-bulkfield}) one finds
\begin{align}\label{fff}
f(R(r,y=0))= C_2r^{n}\left(\frac{f_3}{r^2}+\frac{2\hat{N}^2}{{C^4}}-\frac{4\hat{\hat{N}}}{C^2}\right),
\end{align}
where $f_{3}=\frac{4}{{\beta}}[- \sqrt{\alpha^2+10\alpha+1}-\beta-\alpha^2+2\alpha+2]$.

At $y=0$, one may use the bulk metric to find the Ricci scalar as
\begin{align}
R=\frac{1}{C^2}\left(\frac{\hat{N}^2}{2C^2}-2\hat{\hat{N}}\right)+\frac{2}{\beta r^2}(\beta-\alpha^2-\alpha-1).
\end{align}
It is clear that one may substitute $r$ in terms of $R$ in equation (\ref{fff})
to derive the action on the brane. However, the same is not possible for the bulk
action. This is due to the fact that we have derived all  equations on the
brane by setting $y=0$ and therefore there remains no information
on the extra dimensional dependency.
As a result, starting from a given metric and using the observed physical properties of a particular galaxy such as $v_{tg}$, we may derive the form of $f(R)$ on the brane for the region of flat rotation curves of the given galaxy. Such a $f(R)$ function may now be used for the flat rotation curves of any other galaxy to find the physical parameters and hence the metric.

%***************************************************************************************************************************************
\section{Light deflection angle}
\begin{figure}
\centering
\includegraphics[scale=1]{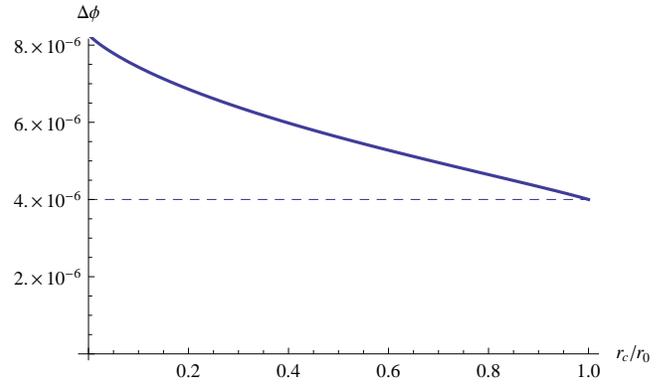}
\caption{Variation of the light deflection angle as a function of $r_c/{r_0}$ for $v_{tg}=300km/s$.}\label{FIG2}
\end{figure}

To calculate the light deflection angle we use the metric found in the region of flat rotation curves, in the form given by (\ref{line1}).
The bending of light results in a deflection angle $\Delta \phi$ given by
\begin{align}\label{141}
\Delta \phi =2 \mid \phi(r_c)-\phi_{\infty}  \mid -\pi,
\end{align}
where $\phi_\infty$ is the incident direction and $r_c$ is the coordinate radius of the closest approach to the center of the galaxy  \cite{A222-9}. Using the geodesic equations, deflection of light can be written as
\begin{align}\label{142}
\phi(r_c)-\phi_\infty =\int_{r_c}^\infty {e^{\frac{\lambda(r)}{2}}\left[ e^{\nu(r_c)-\nu(r)}\left(\frac{r}{r_c}\right)^2 -1\right]^{-\frac{1}{2}}}\frac{dr}{r}.
\end{align}
We consider the case where $r_c$ is in the region of  flat rotation curves. Therefor, the integral is split into two parts
\begin{align}\label{142.5}
&\phi(r_c)-\phi_\infty =\nonumber \\
&\int_{r_c}^{r_0} \frac{B}{k} (1-v_{tg}^2)\left[\left(\frac{r_c}{r}\right)^{(2v_{tg}^2-2)}-1  \right]^{-1/2}\frac{dr}{r} \nonumber\\
&+\int_{r_0}^ \infty \frac{1}{\sqrt{1-\frac{2G_NM}{r}}}\left[\frac{1-\frac{2G_NM}{r_c}}{1-\frac{2G_NM}{r}} \left(\frac{r}{r_c}\right)^{2}-1\right]^{-1/2} \frac{dr}{r},
\end{align}
where $r_0$ is the radius for which the contribution of dark matter vanishes. The first integral is attributed to the region of flat rotation curves with the metric given by equation (\ref{line1}) where dark matter is dominant. The second integral relates to the exterior region with the Schwarzschild metric \cite{A222-9}. As a result we  obtain
\begin{align}\label{142.6}
\Delta\phi= & \bigg |  \frac{2}{\sqrt{1-\frac{2G_NM_b}{R}}(v_{tg}^2-1)} \bigg[\arctan{\frac{1}{{\sqrt{\left(\frac{r_c}{r_0}\right)^{(2v_{tg}^2-2)}-1}}}}\nonumber\\
&-\frac{\pi}{2}\bigg] +2\arcsin {\left(\frac{r_c}{r_0}\right)}-\frac{2G_NM}{r_c}\bigg[ -2+\sqrt{1-\left(\frac{r_c}{r_0}\right)^2}\nonumber \\
&+\sqrt{\frac{r_0-r_c}{r_0+r_c}} \bigg] \bigg | - \pi,
\end{align}
as the total light deflection angle.
It is clear that in the case $R=r_0=r_c$, the deflection angle only originates from baryonic matter. In this particular case, we have $\Delta \phi=4 G_N M_b/R=4 \times 10^{-6}$ which is consistent with the general relativity prediction. The effects of dark matter on the deflection angle is shown in Fig.\ref{FIG2} where variation of $\Delta \phi$ as a function of $r_c/{r_0}$ is plotted.
It is necessary to point out that this result is valid for the region $R < r_c < r_0$. In general, the point of the closest
approach, $r_c$, can be anywhere.
%IIIIIIIIIIIIIIIIIIIIIIIIIIIIIIIIIIIIIIIIIIIIIIIIIIIIIIIIIIIIIIIIIIIIIIIIIIIIIIIIIIIIIIIIIIIIIIIIIIIIIIIIIIIIIIII
\section{Radar echo delay}
The time needed for light to travel by a massive celestial body is longer than the same time calculated by Newtonian gravity. This effect originates from the existence of gravitational field in the vicinity of these massive objects and can be detected using a radar echo delay. Such a  delay was measured by a method suggested by Shapiro \cite{Shapiro}. The time needed for a photon to propagate from one point with $r=r_1$, $\phi=\phi_1$, $\theta=\frac{\pi}{2}$ to another point with $r=r_2$, $\phi=\phi_2$, $\theta=\frac{\pi}{2}$ can be obtained by use of geodesic equations  \cite{A222-9}. The elapsed time for a photon  traveling between $r_c$ and $r$ is given by
\begin{align}
t(r,r_c)=\int_{r_c}^r\frac{ e^{\lambda(r)/2}}{e^{\nu(r)/2}}\left[1-\frac{e^{\nu(r)}}{e^{\nu(r_c)}}\left(\frac{r_c}{r}\right)^2 \right]^{\frac{-1}{2}}dr, \label{time}
\end{align}
where $r_c$ is the closest approach to the celestial body.

To find the time delay we suppose that $r$ is outside the region formed by the galaxy and $r_c$ is in the region of flat rotation curves. Thus, we split the integral (\ref{time}) into two parts according to
\begin{align}
&t(r,r_c)=\nonumber \\
&\int_{r_c}^{r_0} \frac{B}{Ck} (1-v_{tg}^2)\left[r^{2v_{tg}^2}\left(1-\left(\frac{r}{r_c}\right)^{(2v_{tg}^2-2)}\right)  \right]^{-1/2}dr \nonumber\\
&+\int_{r_0}^ r \frac{1}{1-\frac{2G_NM}{r}}\left[1-\frac{1-\frac{2G_NM}{r}}{1-\frac{2G_NM}{r_c}} \left(\frac{r_c}{r}\right)^{2}\right]^{-1/2} \frac{dr}{r},
\end{align}
where we can integrate the first term exactly and the second term by using the Robertson expansion \cite{A222-9} to find
\begin{align}
t&(r,r_c)\approx\frac{B}{Ckr_0^{v_{tg}^2-1}}\sqrt{1-\left(\frac{r_0}{r_c}\right)^{2v_{tg}^2-2}}+\sqrt{r^2-r_0^2}\nonumber \\
&+2G_NM \ln\left(\frac{r+\sqrt{r^2-r_0^2}}{r_0}\right)+G_NM\left(\frac{r-r_0}{r+r_0}\right)^{1/2}.
\end{align}
The term $\sqrt{r^2-r_0^2}$ is what we would expect if light traveled in a straight line with unit velocity. The first term in the above expression is the effect of the flat rotation curves region and the other terms represent the effect of  general relativity. It is clear that in the absence of the region containing the flat rotation curves, $r_c=r_0$, one gets the time delay for general relativity only.

\section{Conclusions}
In this work we have considered a brane-$f(R)$ model as a possible candidate to explain the question of dark matter. This was achieved by considering the field equations on the brane obtained through the projection of the field equations in the bulk together with the assumption that our spacetime admits a family of conformal symmetries. The additional terms thus appearing in this process in the field equations can be considered as a source for dark matter. We used this  procedure to explain the problem of what is known as the flat rotation curves and obtained an expression for the tangential velocity of a test particle moving in such a region, consistent with  present observational data. In addition, the angle representing the deflection of light and the time representing the radar echo delay of  photons passing through such a region were calculated.

To look further afield, one may be interested to investigate the stability of the model and its consistency with Parameterized Post-Newtonian (PPN) parameters. Four dimensional $f(R)$ gravity models have been considered before in order to study their stability \cite{stability1,stability2} and  PPN parameters \cite{PPN1,PPN2}. This is basically done by investigating weak field limit of the theory involved. The same route can be taken here, leading to the possibility of direct comparison of predictions of the model with observational data.
The weak field limit would also afford the possibility of obtaining the potential energy which is needed to find the Tully-Fisher relation \cite{Tully1,Tully2}.
This important relation  establishes a connection between the rotational velocity of a spiral galaxy  and its luminosity. Since the luminosity is proportional to the mass of the galaxy, the Tully-Fisher relation connects the rotational velocity of a galaxy with its mass. In addition, the virial theorem provides a relation between the kinetic energy (velocity) and potential energy (relating to mass or
luminosity) in self-gravitating systems in equilibrium \cite{ketab2}. As a result, the potential energy found from the weak field limit equations accompanied by the Newtonian virial theorem \cite{Tully1} points to a possible derivation of the Tully-Fisher relation in the present context which furnishes another possibility for investigating the validity of the model.

\end{document}